\documentclass[aps,prb,twocolumn,10pt,showpacs,superscriptaddress,longbibliography]{revtex4-2}
\usepackage[unicode=true,pdfusetitle,
 bookmarks=true,bookmarksnumbered=false,bookmarksopen=false,
 breaklinks=false,pdfborder={0 0 1},backref=false,colorlinks=false]
 {hyperref}
\hypersetup{colorlinks,linkcolor=blue,citecolor=blue,urlcolor=blue}

\usepackage{amsmath,amssymb}
\usepackage{graphics}
\usepackage{tikz}
\usepackage{xcolor}
\usepackage{braket}
\usepackage{ulem}

\newcommand{\mH}{\mathcal{H}}
\newcommand{\bfS}{\mathbf{S}}
\newcommand{\bfk}{\mathbf{k}}
\newcommand{\bfq}{\mathbf{q}}
\newcommand{\bfr}{\mathbf{r}}
\newcommand{\mJ}{\mathcal{J}}
\newcommand{\Cv}{C_{\rm v}}
\newcommand{\triline}{
    \vspace{0.0em}
    \begin{tikzpicture}[x=0.05in,y=0.05in,line width=1]
        \draw (0,0) -- (0,1);
        \draw (0,0) -- (0.866,-0.5);
        \draw (0,0) -- (-0.866,-0.5);
    \end{tikzpicture}
}
\newcommand{\trilineflip}{
    \vspace{0.0em}
    \begin{tikzpicture}[x=0.05in,y=0.05in,line width=1]
        \draw [rotate=180] (0,0) -- (0,1) -- (0,0) -- (0.866,-0.5) -- (0,0) -- (-0.866,-0.5);
    \end{tikzpicture}
}
\newcommand{\mytriangle}{
    \vspace{0.0em}
    \begin{tikzpicture}[x=0.04in,y=0.04in,line width=0.4]
        \draw [densely dotted, cap=round] (0,1.0) -- (0.866,-0.5) -- (-0.866,-0.5) -- (0,1.0);
    \end{tikzpicture}
}

\newcommand{\bfSt}{\mathbf{S}_{\mytriangle}}
\newcommand{\TMPX}{TMPX$_3$}

\begin{document}

\title{Spiral-spin-liquid behaviors and persistent reciprocal kagom\'{e} structure \\
in frustrated van der Waals magnets and beyond}

\author{Chun-Jiong Huang}
\affiliation{Department of Physics and HKU-UCAS Joint Institute 
for Theoretical and Computational Physics at Hong Kong, 
The University of Hong Kong, Hong Kong, China}
\author{Jian Qiao Liu}
\affiliation{State Key Laboratory of Surface Physics and
Department of Physics, Fudan University, Shanghai 200433, China}
\affiliation{Department of Physics and HKU-UCAS Joint Institute 
for Theoretical and Computational Physics at Hong Kong, 
The University of Hong Kong, Hong Kong, China}
\author{Gang Chen}
\email{gangchen@hku.hk}
\affiliation{Department of Physics and HKU-UCAS Joint Institute 
for Theoretical and Computational Physics at Hong Kong, 
The University of Hong Kong, Hong Kong, China}

\date{\today}

\begin{abstract}
We study classical $J_1$-$J_2$ models with distinct spin degrees of 
freedom on a honeycomb lattice. For the XY and Heisenberg spins,
the system develops a spiral spin liquid (SSL) that is a thermal 
cooperative paramagnetic regime with spins fluctuating around 
the spiral contours in the momentum space, and at low temperatures 
supports a vector spin-chirality order despite the absence of 
long-range magnetic order. In a strong contrast, for the Ising moments, 
the low-temperature spin correlation forms a reciprocal 
``kagom\'e'' structure in the momentum space that resembles 
the SSL behaviors and persists for a range of exchange couplings. The 
unexpected emergence and persistence of the reciprocal ``kagom\'e'' structure
 are attributed to the stiffness of the Ising moments and the frustration.
At higher temperatures when the thermal fluctuations are strong 
and the spin correlation is not fully melted, 
the reciprocal structures evolve from ``kagom\'e'' structure
towards the ones demanded by the soft-spin limit. 
This contrasts strongly with the behaviors of the spiral contours 
in the SSL regime for the continuous spins. 
We suggest various experimentally relevant systems including 
van der Waals magnets such as the transition-metal 
phosphorus trichalcogenides TMPX$_3$, Cr$_2$Ge$_2$Te$_6$, the
rare-earth chalcohalides (such as HoOF, ErOF and DyOF) and other 
isostructural systems to realize the SSL-like behaviors and/or the 
reciprocal kagom\'{e} structure. 
\end{abstract}

\maketitle

\section{Introduction}
\label{sec:intro}

% comment on self-consistent gaussian approximation...

% largely expanded to include prediction of specific material.

The recently developed van der Waals (vdW) materials 
provide an excellent platform for the understanding of the 
two-dimensional physics and the potential application of 
various devices~\cite{Gong2019two,Park2016,
Wang2018new,Burch2018magnetism,Yahya2020two}. 
The vdW materials are three dimensional but due to the 
 weak van-der-Waals force between the adjacent layers, 
 a monolayer of vdW materials can be obtained through 
 various exfoliation methods~\cite{Yahya2020two}. 
 The transition-metal phosphorus trichalcogenides, \TMPX, 
 are a class of vdW materials where the transition metals (TM) 
 are combined with phosphorus (P), and chalcogenides (X = S, Se, Te). 
 In such materials, TMs constitute a honeycomb lattice 
 with intrinsic magnetism, and most members of the family 
 exhibit an antiferromagnetic exchange~\cite{Chittari2016electronic}.
Apart from their excellent structure, the magnetic vdW materials 
own their advantage in terms of their variety and the controllability. 
The spin Hamiltonian differs from material to material. 
The spin degrees of freedom in these materials could be of the Ising, 
XY or Heisenberg types~\cite{Joy1992magnetism,
Wildes2015magnetic,Gong2019two}. 
For some materials, e.g. Cr$_2$Ge$_2$Te$_6$, the spin type can even 
be tuned by applying the hydrostatic pressure~\cite{Lin2018,Sun2018effects,Sakurai2021}. 
Moreover, it is convenient to vary the anisotropic interactions 
by the external perturbations, such as the gating and strain, 
or the proximity effects~\cite{Sakurai2021,Gong2019two}.
As far as we are aware, in application, most of the current efforts are 
devoted to the realization of the long-range magnetic orders on the vdW 
materials as the magnetic orders are partially forbidden by 
the Mermin-Wagner theorem, and this may be used in 
designing magnetic devices. Other efforts have been devoted 
to exploring the interesting magnetic excitations, 
such as the topological magnons~\cite{Weyl2016,mcclarty2021topological} 
for the honeycomb lattice antiferromagnet CrI$_3$, 
with respect to the magnetically ordered 
ground states~\cite{PhysRevX.8.041028,liu2020observation}.

Here, we deviate from the practical purpose of device designing 
with vdW magnets, and instead address the possibility of interesting 
fundamental physics that could potentially occur in these new materials. 
%The physical mechanism for interesting fundamental physics may bring unexpected possibility of application. 
The direction that 
we are toward here is magnetic frustration. 
Frustrated magnetism has attracted tremendous interest 
for decades because of many unconventional and exotic 
properties~\cite{Savary2016quantum,Broholm2020quantum} 
and the potential application to quantum computing and quantum information.     
Under thermal or quantum fluctuations or both, exotic states, 
i.e. classical or quantum spin liquids~\cite{Savary2016quantum}, 
spin ice~\cite{Henley2010the,Hermele2004pyrochlore,Gingras2014quantum}, 
Kitaev spin liquid~\cite{Kitaev2006anyons}, 
Berezinskii-Kosterlitz-Thouless (BKT) phase~\cite{Isakov2003interplay} 
and many others could emerge. 
Frustration can come from the lattices themselves.
The well-known lattices of geometric frustration are the triangular lattice, 
the kagom\'e lattice, and the pyrochlore lattice.
Generally speaking, frustrations come from the competing interactions, 
e.g. the competition between the inter-sublattice and the intra-sublattice 
interactions. The antiferromagnetic $J_1$-$J_2$ spin model on a square lattice 
is a simple example of competing interactions~\cite{Chandra1988possible}. 
Another well-known example is the same model 
but on a diamond lattice~\cite{Bergman2007order}. 
In this model, there exists a spiral spin liquid (SSL) regime within some parameter region 
and this physics is already detected in experiments in the diamond
lattice antiferromagnet MnSc$_2$S$_4$~\cite{Gao2017spiral}.
The SSL is a special family of classical spin 
liquids~\cite{Moessner1998properties,Castelnovo2008,Henley2010the,
Gingras2014quantum,Wills2002model,Moessner2003theory,
Isakov2004magnetization,Wannier1950antiferromagnetism,
Baskaran2008spin,Price2012critical,Price2013finite,Bergman2007order}, 
and it exhibits partial degeneracies where, in the thermodynamic limit, 
the spin structure factor displays the spiral surfaces or spiral contours 
in the reciprocal space~\cite{Bergman2007order,
PhysRevB.96.085145,Mulder2010spiral,
PhysRevLett.117.167201,Buessen2018quantum,Niggemann2019classical,
pohle2017spin,pohle2021theory,Yao2020generic,PhysRevB.97.115102,JPSJ.79.114705}. 
To our knowledge, there are relatively limited numbers of works about the SSL physics, 
and most previous studies have focused on the Heisenberg spins. 
The two-dimensional (2D) vdW magnets provide an opportunity to explore the 
physics of the SSLs and render new magnetic degrees  of 
freedom beyond the Heisenberg spins for the study of the SSLs
and other frustrated spin physics.

Apart from these transition metal-based-vdW magnets, 
there has been intensive interest in the rare-earth magnets.
Recently, the rare-earth honeycomb lattice magnets have been 
proposed as candidate for Kitaev materials~\cite{Li_2017,2019PhRvB..99x1106J,Luo_2020}, 
and a series of vdW rare-earth chalcohalides with an equivalent 
honeycomb geometry have been synthesized~\cite{2021arXiv210312309J}. 
Anisotropic interactions are quite common 
for the rare-earth magnets~\cite{Luo_2020}, 
and one such anisotropic limit is the Ising
model when the spin moment is Ising-like. 
The Heisenberg model is applicable to the Gd-based magnet 
with ${S=7/2}$ and was also argued to be relevant for some 
Yb-based magnets (where the moment is effective spin-1/2)~\cite{PhysRevB.98.054408}. 
Since many 2D vdW magnets have a 2D honeycomb structure, 
we consider a $J_1$-$J_2$ spin model on a honeycomb lattice with 
the Ising, XY or Heisenberg spins. For the XY and Heisenberg spins,
Refs.~\cite{Bergman2007order,Mulder2010spiral,Niggemann2019classical,
Yao2020generic} have shown that there exist SSLs on bipartite 
lattices such as the honeycomb structure. The magnetic-order transition,
which breaks the U(1) or SO(3) symmetry, will not happen in the 2D 
systems according to the Mermin-Wagner theorem. Nevertheless, 
we find that the $\mathbb{Z}_2$ symmetry, i.e. the chiral symmetry, is 
spontaneously broken at low temperatures when $J_2$ is larger 
than a critical value ${\sim 0.7J_1}$.

Although the continuous spin seems necessary for the construction 
of the spin spirals, a pattern similar to 
the spiral contour for the SSL, a reciprocal ``kagom\'e'' structure,
will emerge for the low-temperature spin correlation in the momentum space 
for the Ising spin. This behavior resembles the SSL for the continuous spin. 
However differing from the varying spiral contour structures with varying parameters, 
our calculation in Sec.~\ref{sec3} shows that this reciprocal ``kagom\'e'' structure 
persists at low temperatures for a range of $J_2$'s. 
 This remarkable result, as we further explain 
in Sec.~\ref{sec3}, is a unique but rather natural property 
of the Ising spin moment and arises from the stiffness of the 
Ising moment and the local constraint due to the frustration 
upon varying of the parameters. 
It is also found interesting to explore the thermal evolution 
of the reciprocal structure. Unlike the robustness against
the variation of $J_2$, the reciprocal structure is found to 
deviate from the reciprocal ``kagom\'e'' as the temperature
is above a crossover temperature in Sec.~\ref{sec4}. 
The reciprocal structure
gradually evolves toward the contours that are demanded by 
the exchange interaction with the given exchange couplings 
in the soft spin limit. This thermal 
behavior at high temperatures is further understood by the soft spins due to 
the thermal fluctuations.

\begin{figure}[t]
    \includegraphics[width=0.85\linewidth]{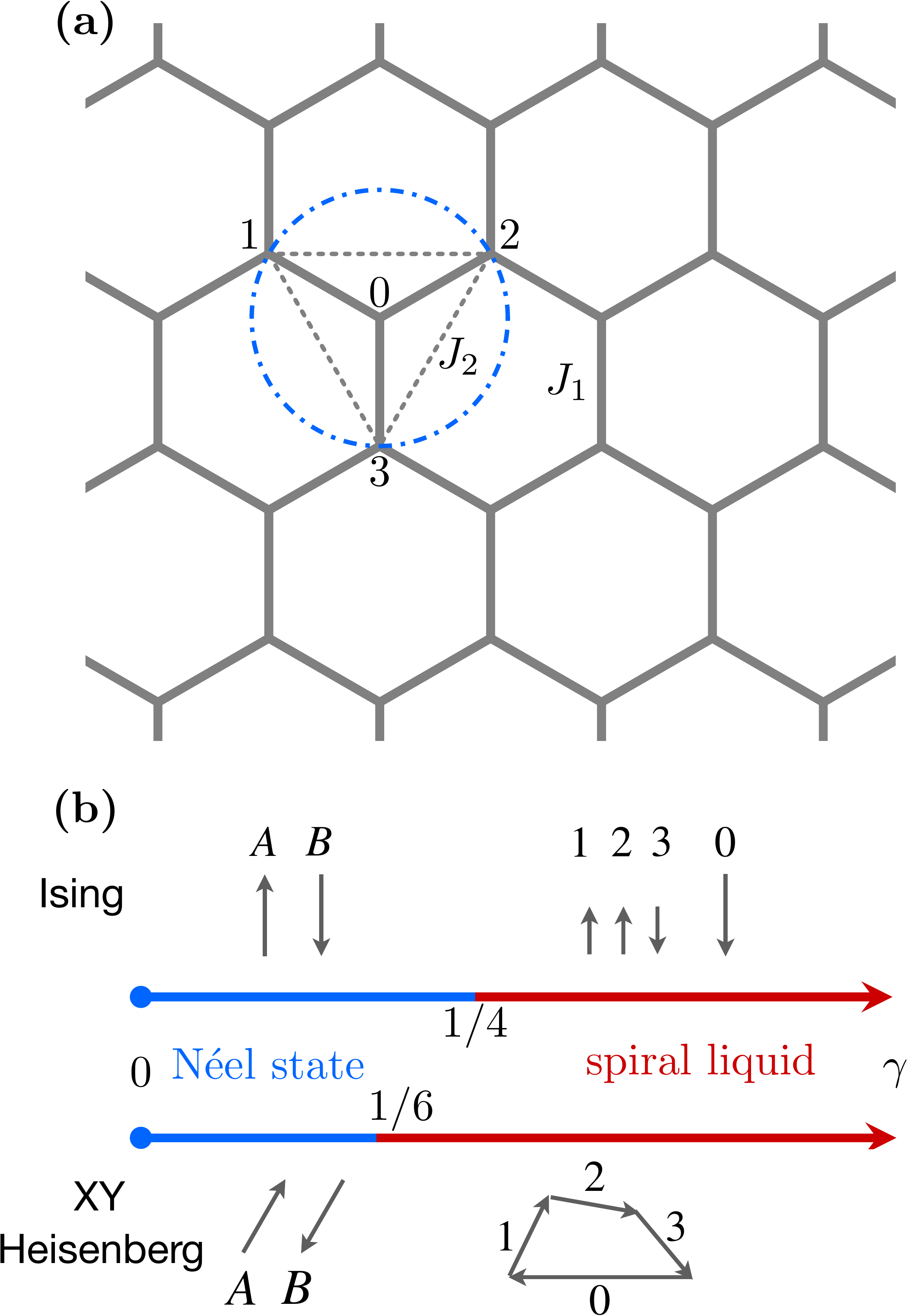}
    \caption{
    (a) Honeycomb lattice with its cluster unit in the circle. 
    (b) Diagram for Ising, XY and Heisenberg spins. 
    For small ${\gamma \equiv J_2/J_1}$, the system is in the N\'eel state, 
    and the spins align oppositely on two sublattices. 
    Massive degeneracies exist when ${\gamma>1/4}$ for the Ising spin 
    where the SSL-like behaviors and other special properties appear at low temperatures,
    and ${\gamma>1/6}$ for the XY and Heisenberg spins 
    where the SSL regime appears at low temperatures. 
    Spins in the cluster are shown as arrows with $\bfS_1,\bfS_2$, and $\bfS_3$ 
     scaled by $2\gamma$. }
    \label{fig1} 
\end{figure}

\begin{figure*}[t]
    \includegraphics[width=1.0\linewidth]{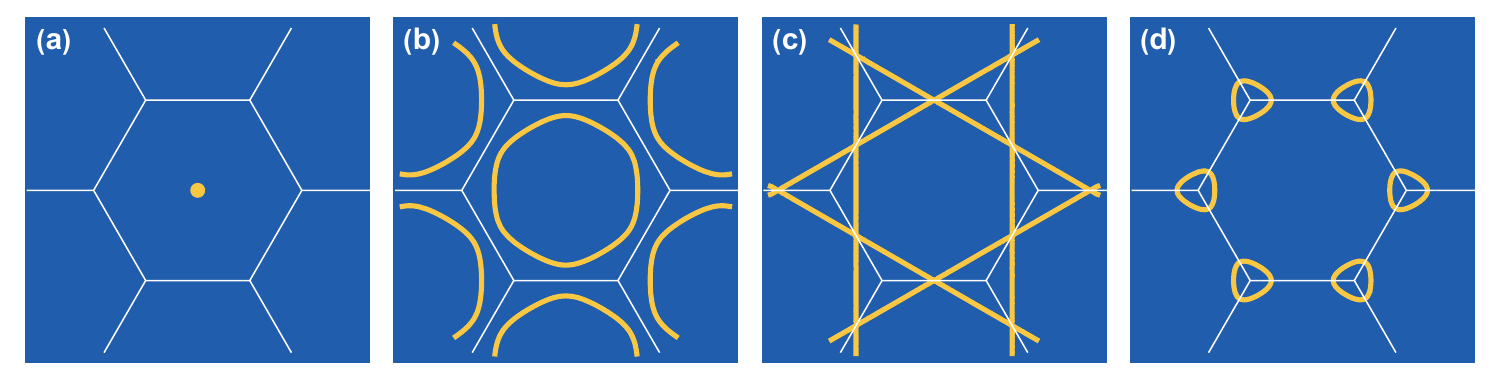}
    \caption{The spin structure factors of the ground state at different $\gamma$'s 
    (${\gamma = J_2/J_1}$) for the XY and Heisenberg spins.
    (a) ${\gamma=0.08}$, (b) ${\gamma=0.40}$, (c) ${\gamma=0.50}$ 
    and (d) ${\gamma=0.80}$. The white hexagon in the center 
    is the boundary of the first Brillouin zone.
   At $\gamma=0.50$ (c), the contours form a reciprocal ``kagom\'{e}'' structure
   in the momentum space.     
    }
\label{fig2}
\end{figure*}

The rest of the paper is organized as follows.
In Sec.~\ref{sec2}, we explain the model and the basic properties
of the SSLs for the continuous spins. Some of the physics is explained
from the local-constraint point of view. While the case of continuous spin 
has been studied in previous work, the perspective of the local energetic 
constraint provides some insights into the emergent properties.
Moreover, we show the appearance of the finite-temperature spin chirality 
order that relates to the local electric polarization via 
the inverse Dzyaloshinskii-Moriya mechanism. In Sec.~\ref{sec3}, we turn 
our attention to the Ising spin where the SSL-like behaviors surprisingly 
occur and persist for a range of parameters at low temperatures. 
Apart from the finite-temperature phase transition 
due to the {\sl discrete} nature of the local moments, the spin correlation supports the
reciprocal ``kagom\'e'' structure in the momentum space at the low temperatures.
This is explained from the local constraint point of view. 
In Sec.~\ref{sec4}, we further explore the temperature evolution of the 
spin correlation for the Ising spin and explain the peculiar crossover behaviors
in the reciprocal structures. 
A comparison with the self-consistent Gaussian approximation is also given. 
Finally in Sec.~\ref{sec5}, we summarize our results
 and discuss the experimental relevance. We particularly emphasize 
 the Ising spin connection with the rare-earth chalcohalides 
 (such as HoOF, ErOF and DyOF) and make predictions based on 
 our theoretical results.

\section{Model with continuous spins}
\label{sec2}

We consider the $J_1$-$J_2$ spin model on the honeycomb lattice with periodic
boundary conditions (Fig.~\ref{fig1}),
\begin{equation}
    \mH = J_1 \sum_{\left< ij \right>} \bfS_i \cdot \bfS_j + %
          J_2 \sum_{\left[ ij \right]} \bfS_i \cdot \bfS_j,
\label{eq:h}
\end{equation}
where $\left< ij \right>$ ($\left[ ij \right]$) refers to the nearest (next-nearest)
neighbors and $\bfS_i$ is the classical Ising, XY or Heisenberg spin on
site $i$. The exchange coupling $J_2$ is antiferromagnetic, and $J_1$ 
could be ferromagnetic or antiferromagnetic because a simple 
transformation on one sublattice switches the sign of $J_1$. 
Here we set $J_1$ to be antiferromagnetic. With only $J_1$, 
the ground state is a simple N\'eel state because of the bipartite 
nature of the honeycomb lattice. With only $J_2$ exchange, 
the Hamiltonian decouples to two independently antiferromagnetic 
spin models on two triangular sublattices. For the Ising spins, 
the ground state is massively degenerate, 
and it is $120^\circ$ state for the XY and Heisenberg spins.
When $J_1$ and $J_2$ are both non-zero, a large frustration is introduced 
when $J_2$ is larger than a critical value $J_{2c}$ and interesting properties 
could appear due to the strong frustration. To obtain comprehensive 
and accurate behaviors of this model, we mainly employ the classical 
Monte Carlo (MC) method to investigate both the zero- and 
finite-temperature properties for three different types of 
the spin moments.

\subsection{Analytical results for continuous spins}
\label{sec2a}

In this section, we focus on the XY and Heisenberg spins for 
which the model has a global U(1) and SO(3) symmetry, respectively.  
Thus, there is no finite-temperature magnetic ordering transition 
according to the Mermin-Wagner theorem. Although part of the results 
in this section was previously known in Ref.~\cite{Mulder2010spiral},
which studied the quantum Heisenberg spins, 
we include this analysis for completeness and 
for later comparison with the peculiar Ising case. 
In the Luttinger-Tisza method, 
the local constraint ${|\bfS_i|^2=S^2}$ for each spin is softened to 
a weak global constraint ${\sum_i |\bfS_i|^2 = N S^2}$, 
where $N$ is the number of lattice sites. 
In this weak constraint, we minimize the energy
and check whether the strong constraints are satisfied afterwards.
 If these strong constraints are satisfied as well, 
 the ground state obtained from the Luttinger-Tisza method 
 is just the ground state of the initial model.
In practice, we define the Fourier transformation 
of $\bfS_i$ in the sublattice ${\mu = A, B}$ as
${\bfS_i^\mu={(N/2)^{-1/2}}\sum_\bfk e^{i\bfk\cdot \bfr_i} \bfS^\mu(\bfk)}$. 
The Hamiltonian in Eq.~\eqref{eq:h} can be rewritten as 
\begin{equation}
\mathcal{H}=\frac{1}{2} 
\sum_{\bf k}    
\phi (-{\bf k})^T \mJ({\bf k})\, \phi (\bf k),
\end{equation}
with ${\phi(\bfk)=(\bfS_A({\bf k}), \bfS_B({\bf k}))^T}$. 
Here $\mJ(\bfk)$ is the Fourier transform of the
adjacency matrix on the honeycomb lattice and is given as 
\begin{equation}
\label{eq:LTmatrix}
    \mJ(\bfk) = \begin{bmatrix}
    J_2 [\Lambda(\bfk)^2-3]
    & J_1 \sum_{\mu} e^{i\bfk \cdot  \mathbf{b}_\mu} \vspace{3mm}\\
    J_1 \sum_{\mu} e^{-i\bfk \cdot \mathbf{b}_\mu}
    &J_2 [\Lambda(\bfk)^2-3]
    \end{bmatrix} ,
\end{equation}
where 
 the vectors $\mathbf{b}_\mu$ refer to the nearest neighbor vectors of the
honeycomb lattice. The minimal eigenvalue of matrix $\mJ(\bfk)$ is given as
\begin{equation}
\label{eq:minE}
      \epsilon_{-} ({\bfk}) = 
     J_2 \left(\Lambda(\bfk)-\frac{J_1}{2J_2}\right)^2 
     - \frac{J_1^2+12J_2^2}{4J_2},
\end{equation}
with 
\begin{equation}
{\Lambda({\bfk})  =\Big|\sum_{\mu} e^{-i{\bfk}\cdot \mathbf{b}_\mu}\Big|} ,
\end{equation}
and the energy of the ground state is the minimum value of $\epsilon_{-} ({\bfk})/2$.
The range of $\Lambda({\bfk})$ is from 0 to 3, and ${\gamma_c=(J_2/J_1)_c=1/6}$ 
is a critical point. When ${\gamma\equiv J_2/J_1 <\gamma_c}$, the point of 
${\Lambda({\bf q})=3}$ will minimize $\epsilon_-({\bf q})$, 
i.e., ${\bf q}=0$, and this corresponds to the N\'eel order. 
For larger values of $\gamma$, $\epsilon_-({\bf q})$ takes the minimum when 
${\Lambda(\bfq)=(2\gamma})^{-1}$, and these momentum vectors constitute 
closed contours in the reciprocal space, dubbed the spiral contours, as shown 
in Fig.~\ref{fig2}. Clearly, the system has a massively degenerate ground-state
manifold, indicating a strong frustration. 
With increasing $\gamma$, the spiral contour expands
around the $\Gamma$ point and touches the $M$ point 
when ${\gamma=1/2}$. For a larger ${\gamma>1/2}$,  
a single spiral contour splits into several contours around 
the $K$ points in the first Brillouin zone and they 
gradually shrink to $K$ points. At the limit ${\gamma\rightarrow\infty}$ 
or ${J_1=0}$, the spiral contours disappear leaving only a single spiral state, 
the $120^\circ$ state. 
The model reduces to an antiferromagnetic model on the triangular lattice.

\begin{figure*}[t]
    \includegraphics[width=1.0\textwidth]{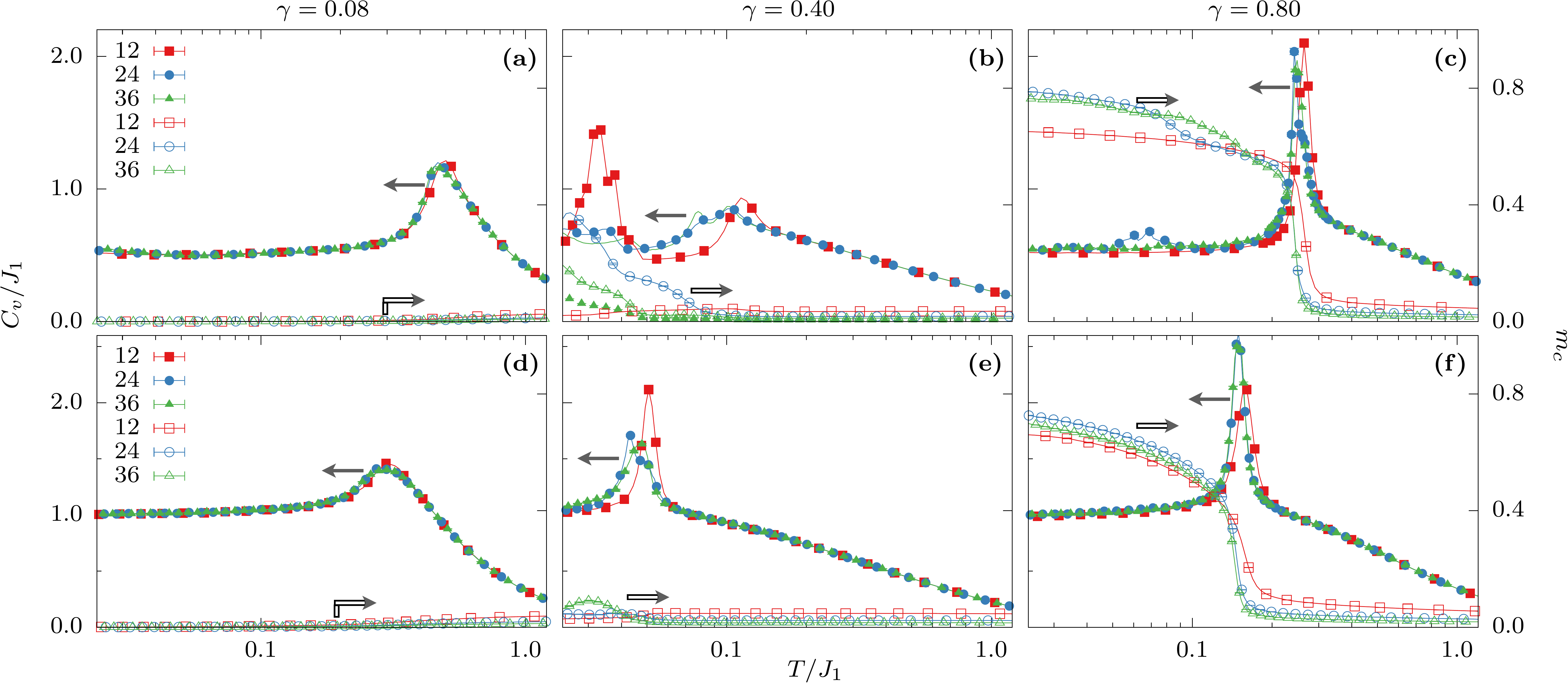}
    \caption{Specific heats $\Cv$ and chiral order 
    parameter $m_c$ for the XY and Heisenberg spins
    at different $\gamma$'s.         
    The lines with solid symbols are the specific heat curves and the 
    lines with open symbols are the chiral order parameter curves. 
    (a)-(c) Plots for the XY spin. 
    (d)-(f) Plots for the Heisenberg spin.
    All curves are plotted with one standard error.}
    \label{fig3}
\end{figure*}

An intuitive method of obtaining the information about the ground state 
is a geometric one. This method is often adopted in frustrated 
systems to find a local constraint to minimize the classical
(and sometimes even quantum) ground-state energy. 
In the classical spin ice on a pyrochlore lattice, for example, the ice rule 
is a local constraint that should be satisfied in the ground-state 
manifold~\cite{Moessner1998properties}. The key ingredient of this trick 
is to split the Hamiltonian into equivalent cluster units, and
then one minimizes the energy of each cluster unit to obtain the lowest energy and 
acquire the local constraints for each cluster unit. We rewrite Eq.~\eqref{eq:h} as
\begin{equation}
    \label{eq:hc}
    \begin{split}
    \mH &= \frac{J_1^2}{8J_2}\sum_{\triline} \left[ \big(\bfS_0
     +\frac{2J_2}{J_1}\bfSt \big)^2 - \frac{J_1^2+12J_2^2}{J_1^2} \right]
      \\
        &= \frac{J_1^2}{8J_2}\sum_{\triline} \left[ 
        \left(\bfS_0+2\gamma\bfSt\right)^2
               - (1+12\gamma^2) \right],
    \end{split} 
\end{equation}
where ${\bfSt=\bfS_1+\bfS_2+\bfS_3}$ is the sum of three corner spins 
on a unit $\triline$/$\trilineflip$ and $\bfS_0$ is the central spin on the unit. 
In the following, we will use $\triline$ to represent both $\triline$ and $\trilineflip$. 
Figure~\ref{fig1}(a) shows the honeycomb lattice and a unit. 
In a honeycomb lattice with $N$ sites, 
there are $N$ $\triline$'s. From Eq.~\eqref{eq:hc} 
one sees that ${\gamma_c=1/6}$ is a critical value 
and is the same as the Luttinger-Tisza result because the largest 
length of $\bfSt$ is three times the length of $\bfS_0$.
If ${\gamma<\gamma_c}$, the minimal energy is reached 
when ${\bfSt=-3\bfS_0}$, i.e. $\bfS_1, \bfS_2$, and $\bfS_3$ are 
all antiparallel to $\bfS_0$. This state is simply the N\'eel state.
In the region of ${\gamma > \gamma_c}$, the condition,
\begin{equation}
\bfS_0+2\gamma\bfSt=0 ,
\label{eq7}
\end{equation} 
for each unit $\triline$ can always be satisfied for the XY and 
Heisenberg spins to minimize the energy. In Fig.~\ref{fig1}(b), 
we depict the evolution of the spins on $\triline$ 
with increasing $\gamma$. We find that there exists
a degeneracy of each unit $\triline$ under the 
local constraint ${\bfS_0+2\gamma\bfSt=0}$ when 
${\gamma \ge \gamma_c}$. This leads to the massive 
degeneracy of the ground state. In the limit ${\gamma=\infty}$ or ${J_1=0}$, 
the constraint on each unit $\triline$ reduces to the weak constraint 
of ${\bfSt=0}$ for each $\mytriangle$ 
and the massive degeneracy is lifted to the discrete $\mathbb{Z}_6$ degeneracy, 
i.e., the ground-state order is the $120^\circ$ state. 
In addition to the continuous rotational symmetry breaking, 
this state also breaks the $\mathbb{Z}_2$ symmetry 
or the chiral symmetry that describes the spin rotation pattern 
on a triangle clockwise or counter-clockwise. 
This symmetry is discrete and can be spontaneously broken 
at finite temperatures, which means that there can be a finite-temperature 
chiral transition under this limit~\cite{Obuchi2012spin,Lv2013phase}.
Moreover, we expect that as long as $\gamma$ is large enough, 
this discrete symmetry can still be spontaneously broken 
at a finite temperature and the chiral order will occur.

The other important quantity is 
the spin structure factor and it can be detected 
experimentally~\cite{Bergman2007order}. 
The low-temperature spin structure factor 
provides an important characterization of the physical properties 
related to the classical ground state degenerate manifold. 
In this paper, we define the spin structure factor on the A or B sublattice 
as 
\begin{equation}
    \label{eq:ssf}
    \begin{split}
    S_{\bfk}^{\mu\mu} &= \frac{1}{N/2}
     \sum_{\bfr_i,\bfr_j} \langle \bfS^\mu_{\bfr_i}\cdot\bfS^\mu_{\bfr_j}  \rangle
     \exp[-i\bfk\cdot(\bfr_i-\bfr_j)] \\
                      &=\langle \bfS^\mu(-\bfk) \bfS^\mu(\bfk)  \rangle,
    \end{split}
\end{equation}
with $\mu=A\text{ or }B$.
In zero temperature, only with those $\bfq$‘s minimized Eq.~\eqref{eq:minE},  
$S_{\bfk}^{\mu\mu}$ are nonzero. The corresponding spin structure factors 
can be expressed as
\begin{equation}
S_{\bfk}^{\mu\mu} = \frac{1}{2}\left( \delta(\bfk-\bfq) + \delta(\bfk+\bfq) \right).
\end{equation}
In Fig.~\ref{fig2}, we plot the spin structure factors for different $\gamma$'s 
where the degenerate momentum vectors form the spiral contours. 
When ${1/6<\gamma<1/2}$, the spiral contour is a single closed loop 
around the $\Gamma$ point in the first Brillouin zone, and its size 
becomes larger with increasing $\gamma$. This contour touches 
the first Brillouin zone boundary at the $M$ point 
when $\gamma$ reaches $1/2$. 
It splits to several contours around the $K$ points when ${\gamma>1/2}$. 
In the limit of ${\gamma\rightarrow\infty}$ or ${J_1=0}$, these spiral 
contours shrink to the $K$ points, which indicates 
the rise of the $120^\circ$ state.

%%%%%

\begin{figure}[t]
    \includegraphics[width=1.0\linewidth]{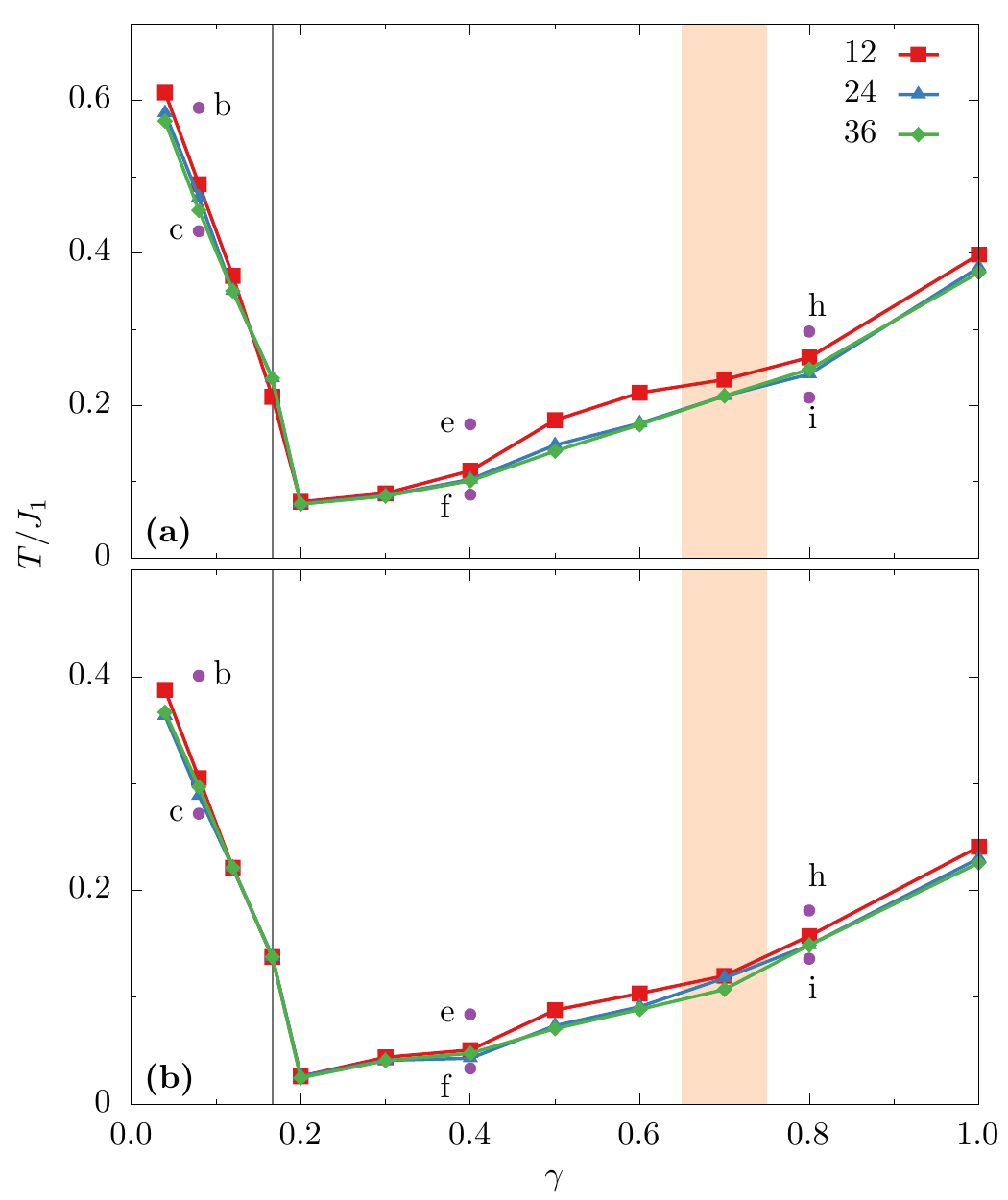}
    \caption{Phase diagrams of (a) XY and (b) Heisenberg spins. 
    Due to the complex behaviors of the specific heat near 
    ${\gamma=0.40}$, the highest-temperature peak is taken 
    as the phase boundary or crossover. 
    Comparing the phase boundaries or transitions of different system sizes, 
    we find that ${L=36}$ is large enough to determine the phase boundaries. 
    The vertical line at ${\gamma_c=1/6}$ separates
    the unfrustrated and frustrated regions. Near ${\gamma=0.70}$, 
    the thermal crossover switches to the phase transition for 
    ${\gamma>0.70}$ where the chiral symmetry 
    is spontaneously broken. 
    This approximate $\gamma$ region is marked with a light orange band.}
       \label{fig4}
\end{figure}

\subsection{Numerical simulation and finite temperature thermodynamics }
\label{subsecb}

Here, we use a Metropolis algorithm combined with the over-relaxation 
method~\cite{Brown1987overrelaxed,Creutz1987overrelaxation} and 
the parallel tempering~\cite{Hukushima1996} to simulate the proposed 
model at finite temperatures. As we are all aware, frustrated 
systems generally have a large energy barrier between numerous local 
minimal energy states with only local updates such as the Metropolis updates. 
This will lead to spin configurations that fluctuate near the minima 
for a long time and make the simulation unreliable. 
To overcome this obstacle, a simple and efficient approach 
is to use the parallel tempering scheme.

In the parallel tempering scheme, multiple replicas of the same system, 
randomly initialized, are simulated at different temperatures. 
Then the exchange of replicas between the nearest temperatures occurs with 
a certain probability, and the exchange-acceptance ratio is calculated 
according to the detailed balance condition. 
Replicas can swap around the whole temperature region 
and this significantly suppresses the configuration freezing 
for replicas at low temperatures. 
This is because updates are efficient for high-temperature replicas 
and these spin configurations can gradually convey to the 
replicas of the low temperatures. The key to this approach 
is to select appropriate temperatures for every replica
to ensure that the exchange-acceptance probabilities between 
the nearest replicas are not too small. In this paper, 
we carefully select the temperatures for each replicas  
to make the exchange probabilities not smaller than 0.32. 
For this goal, we first follow the feedback-optimized plan in 
Ref.~\cite{Hukushima1996} to obtain the tentative temperatures
and the corresponding energies for a small system size. 
According to these temperatures and energies, temperatures 
of replicas for other system sizes can be calculated 
with a given exchanged-acceptance ratio. 
The reason that this strategy works is that the energy density is 
almost independent of the system sizes and the energy-temperature 
curve is continuous in this model. Moreover, we employ the over-relaxation 
method, which could improve the performance of the simulations for the continuous spins.
We carry out 128 independent simulations, and each one contains
multiple replicas of the same system but in different temperatures.
In every independent simulation, a whole MC step consists of a single
Metropolis-update sweep and subsequently a single over-relaxation sweep.
Besides, a parallel-tempering update will occur after every 50 MC steps.
After thermalizing systems to equilibration, ${4\times 10^4}$ samples are 
produced in each simulation, and in total ${5.12\times 10^6}$ samples are 
used for the statistical analysis.

\begin{figure*}[t]
    \includegraphics[width=\textwidth]{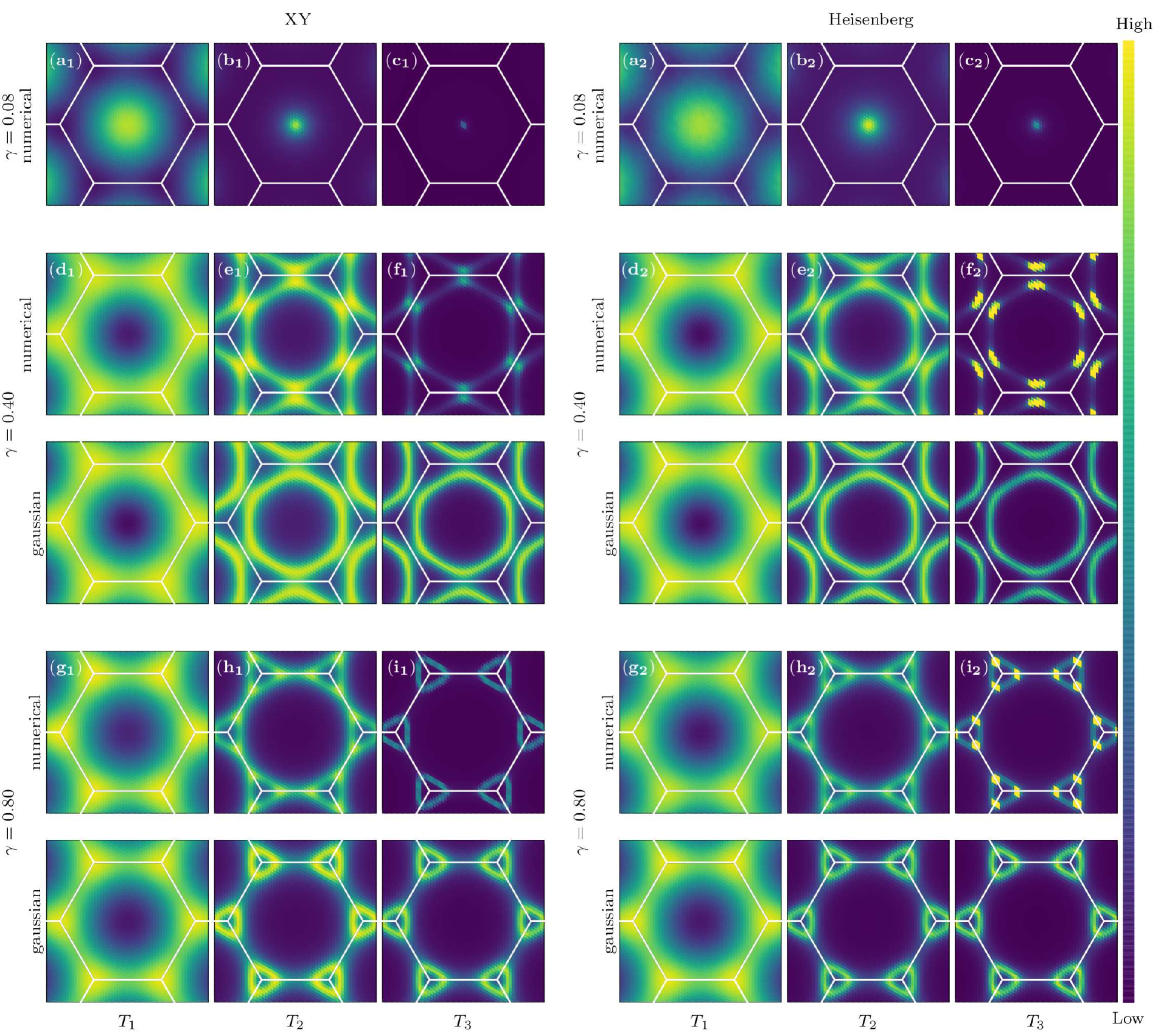}
    \caption{Spin structure factors of the XY spin [left panel,$(a_1)$-$(i_1)$] and
    the Heisenberg spin [right panel,$(a_1)$-$(i_1)$] 
    at ${\gamma=0.08,0.40,0.80}$ with the linear system size ${L=36}$. 
    Temperature ${T_1(3J_1>T_2>T_3)}$, $T_2$, and $T_3$ are marked in 
    Figs.~\ref{fig4}(a) and \ref{fig4}(b) with purple points.
    The intensity bar indicates the relative intensity for each figure.  
    A comparison with the results from the self-consistent
    Gaussian approximation in Sec.~\ref{sec4} is also shown. 
    }
    \label{fig5}
\end{figure*}

To determine the finite-temperature phase diagram, 
we measure the specific heat $\Cv$ and the chiral 
order parameter $m_c$. The peaks of the specific 
heats $\Cv$ can be used to determine the crossovers 
or the phase transition. The chiral order parameter 
$m_c$ on a sublattice lattice 
is defined as
\begin{equation}
    \label{eq:chiral}
    m_c=\frac{1}{N/2} \big|\sum_{t} \sum_{ij\in t} \bfS_i\times\bfS_j \big|,
\end{equation}
where $\sum_{t}$ is taken over all up-triangular units in 
the sublattice A or B and $\sum_{ij\in t}$ sums over 
three bonds of each up triangle $t$ in a clockwise order.
In Figs.~\ref{fig3}(a-c), we plot the curves of the 
specific heat $\Cv$ and the chiral order parameter
$m_c$ at ${\gamma=0.08,0.40,0.80}$ 
with the linear system sizes ${L=12,24,36}$ for the XY spin. 
When ${\gamma=0.08}$, the system undergoes a crossover 
from the quasi-N\'eel state of the low temperatures to 
the paramagnetic state of the high temperatures,
and the specific heat only has a non-divergent round peak 
at the crossover due to the Mermin-Wagner theorem.

After entering the SSL regime for ${\gamma=0.40}$, 
there exists a crossover accompanied by a nondivergent 
round peak without chiral-symmetry breaking.
For a larger next-nearest exchange interaction, ${\gamma=0.80}$, 
the frustration is not large enough to prevent the appearance of 
the spontaneous breaking of the chiral symmetry. 
This leads to a sharp peak in the specific heat but the magnetic order 
is still forbidden by the thermal fluctuation. At the same time, 
the chiral order parameter $m_c$ has a rapid growth near 
the transition indicating the breaking of the chiral symmetry.
Using the highest-temperature peaks of the specific heat 
$\Cv$ and the points of the rapid increase in the chiral 
order parameter $m_c$, the phase diagram of the XY spin can 
be determined, as shown in Fig.~\ref{fig4}(a). 
Near ${\gamma\approx 0.7}$, the crossover changes to a phase transition. 
In the Heisenberg case, the same procedure is applied as shown 
in Fig.~\ref{fig3}(d-f), and Fig.~\ref{fig4}(b) is the phase diagram.
Due to the system size and the numerical method, the potential
BKT transition is not discernible here. 
For both the XY and Heisenberg spins, the behaviors of specific heats
at ${\gamma=0.40}$ are strongly affected by the system sizes. We think this is due to
the stronger frustration. Near ${\gamma=0.25}$,  the energy difference between
the N\'{e}el order and the spiral spin states with the wavevectors from the
spiral contour is small and at finite temperatures,
there exists strong competition between these spin configurations. This 
leads to rather complex behaviors and makes the simulations more difficult.

One outcome of the vector spin chirality order at low temperatures
is the local electric polarization. This is obtained from the
inverse Dzyaloshinskii-Moriya mechanism that gives the 
local electric polarization~\cite{PhysRevLett.95.057205},
${{\boldsymbol P}_{ij} \propto {\boldsymbol e}_{ij} \times 
(  {\boldsymbol S}_i \times {\boldsymbol S}_j  )} $,
where ${\boldsymbol e}_{ij}$ is the unit vector that connects site 
$i$ to site $j$. A finite vector spin chirality order  
implies a distribution of the local electric polarization ${\boldsymbol P}_{ij}$
that could lead to a modification in the electric response
and the structural distortion.

\subsection{Spin structure factors}

Here we further determine the evolution of the spin structure factors at different 
$\gamma$'s for the XY and Heisenberg spins. The spin structure factors can be 
detected by the neutron scattering to reveal the magnetic structures. We adopted 
three different $\gamma$'s and three different temperatures for both spins. 
The results are presented in Fig.~\ref{fig5}.

The temperatures in Fig.~\ref{fig5} are marked in Fig.~\ref{fig4} 
with the purple points except for ${T_1=3J_1}$. We choose 
${T_1 >T_2 > T_3}$ in the plots. In the following, we analyze
the case of the XY spin and it is similar for the Heisenberg spin.
In Fig.~\ref{fig5}(a$_1$-c$_1$) of the XY spin, we set ${\gamma=0.08}$,  
and the peaks of the spin structure factors are always at the $\Gamma$
point as expected for the proximate N\'eel state at zero temperature. 
At the high temperature $T_1$, the system is in the disordered state 
with round peaks; the system is in the quasi-N\'eel state at the low temperature 
$T_3$ with a sharpened peak. In Fig.~\ref{fig5}(d$_1$-f$_1$), we plot the spin 
structure factors for ${\gamma=0.40}$. At $T_1$, there are broad peaks 
located at the $K$ points which means the spin structure factors are 
dominated by the $J_2$ exchange. At temperature $T_2$, the
peaks of the spin structure factors form a visible spiral contour and 
there are massively degenerate states. Further decreasing the temperature to $T_3$, 
the spiral contour becomes more sharp accompanied by some peaks that will
be discussed next. For ${\gamma=0.8}$, the spin structure factors of the 
high temperatures in Fig.~\ref{fig5}(g$_1$) cannot be distinguished from those of 
Fig.~\ref{fig5}(d$_1$). Nevertheless, there are several spiral contours around 
the $K$ points for the low temperatures $T_2$ and $T_3$, 
as shown in Fig.~\ref{fig5}(e$_1$,f$_1$). 
At the temperature $T_3$, the discrete spiral contours are more clear.
For the Heisenberg spin, the numerical results of the spin structure factors are shown
in Fig.~\ref{fig5}(a$_2$-i$_2$).

\begin{figure}[t]
    \includegraphics[width=0.95\linewidth]{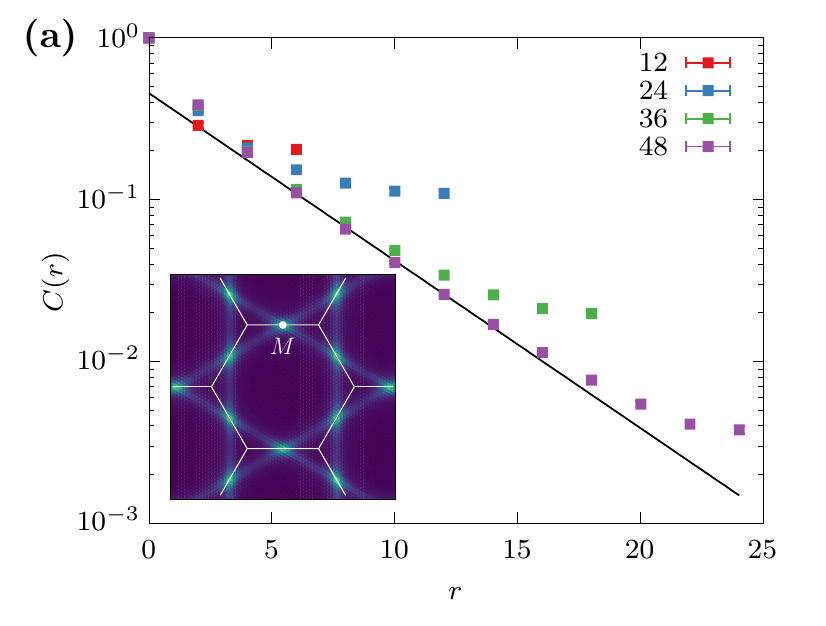}
    \includegraphics[width=0.95\linewidth]{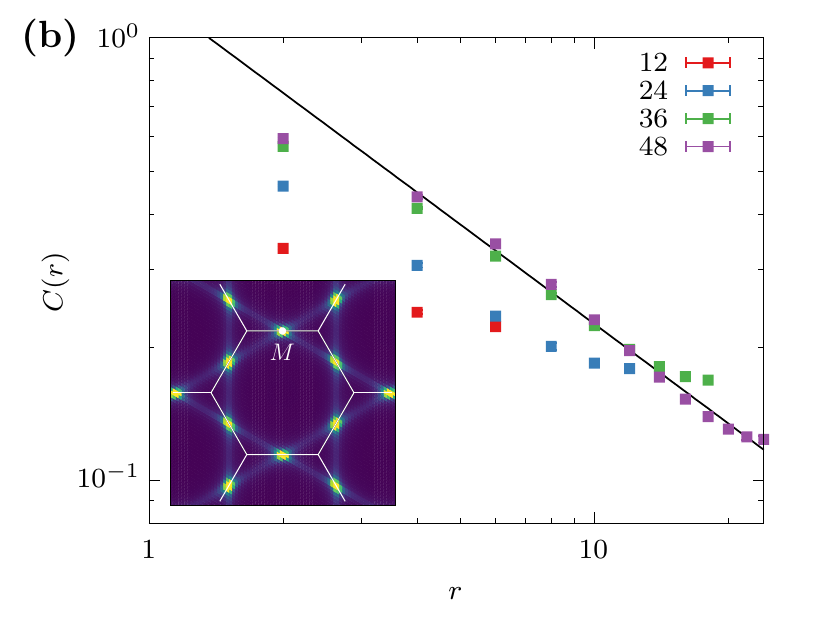}
    \caption{\label{fig6}
    The behavior of the spin correlation $C(r)$ at ${\gamma=0.5}$ for the XY (top) and
    Heisenberg spin (bottom). 
    $C(r)$ decays exponentially with the distance for the XY spin and it is more
    like a power-law function for the Heisenberg spin.
    The inset is the normalized spin structure factor where peaks locate at the $M$ point.
    }
\end{figure}

In Fig.~\ref{fig5}(e$_1$,f$_1$,h$_1$), it seems that 
there are magnetic order peaks. According to the 
Mermin-Wagner theorem, however, this model should 
not have any magnetic order at finite temperatures.
The thermal order-by-disorder mechanism to maximize the entropy work
similarly as the three dimensions~\cite{Bergman2007order,Yao2020generic}. 
Near these wavevectors where
the peaks are located, the entropies are larger than those of remaining parts 
on the spiral contours,  but thery are not strong enough to induce any long-range
magnetic order. With further decreasing temperature, this entropic effect
becomes weaker and ultimately, at zero temperature, the relative difference
disappears as shown in Fig.~\ref{fig2}(b,c) with an equal strength. 
To give a consistency check of the absence of the true long-range order,
we simulate the system at temperatures lower than the crossover temperature for
${\gamma=0.5}$ where the peaks of the spin structure factor are located at the $M$ point.
The $M$ point magnetic order on the triangular lattice was previously
known to be a stripe-type magnetic structure where the spins are ferromagnetically aligned along 
one Bravais lattice vector direction and are antiferromagnetically aligned along 
the other Bravais lattice vector direction~\cite{PhysRevB.94.035107}. 
The results for the spin correlation and the spin structure factors 
are summarized in Fig.~\ref{fig6}.

The spin correlation ${C({\bf r})=\braket{{\bf S}({\bf r}) \cdot {\bf S}(0)}}$ in 
Fig.~\ref{fig6} is measured along one Bravais lattice vector direction 
and the distance $r$ is chosen to be an even lattice spacing. $C(r)$
has no sign oscillation and can reflect whether there is magnetic order. 
In the XY case (Fig.~\ref{fig6}(a)), the spin correlation decays exponentially with increasing
distance.
%and the peaks' strengths at the $M$ point converge to a constant with the system sizes.
%For the Heisenberg spin, the spin structure factors are shown in
%Fig.~\ref{fig5}(a$_2$-i$_2$) and those peaks are also not
%the magnetic order peaks that can be seen from Fig.~\ref{fig6}(b).
%What is different from the XY spin is that the spin correlation $C(r)$ is more
%like a power-law function and the strengths at the $M$ point are divergent with the
%system sizes. This discrepancy makes the peaks of the spin structure factor
%in the Heisenberg spin more pronounced than those in the XY spin as shown in
%Fig.~\ref{fig5}.}
For the Heisenberg spin, $C(r)$ also decays as the distance $r$ increases, as
can be seen from Fig.~\ref{fig6}(b). What is different from the XY spin is that
the spin correlation $C(r)$ is more like a power-law function. This discrepancy
makes the peaks of the spin structure factor in the Heisenberg spin more
pronounced than those in the XY spin as shown in Fig.~\ref{fig6}.
Towards large distances $r$, the deviation from the fitted line is
a consequence of the periodic boundary conditions and, with larger system sizes,
more points will fall on the fitted line.
%{Although the decay behavior of $C(r)$ within the simulated system sizes cannot totally
%rule out the existence of the long-range magnetic order and, considering
%the Mermin-Wagner theorem, we conjecture that there is no magnetic order.}
%However, for other $\gamma$'s, the spatial spin configuration is too complex to
%construct a simple spin correlation in the spatial space.

\begin{figure*}[t!]
    \includegraphics[width=1.0\textwidth]{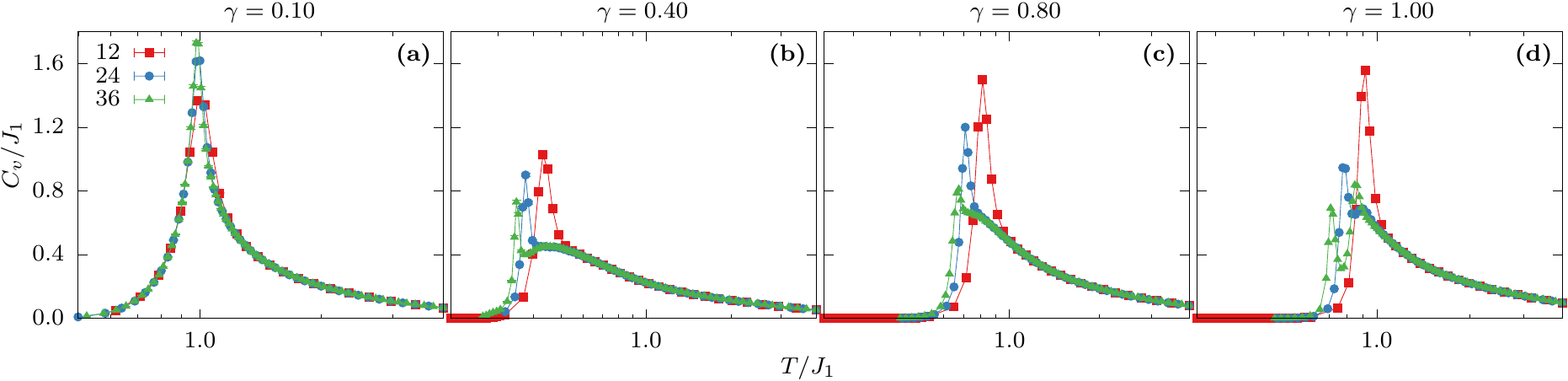}
    \caption{Specific heat $\Cv$ with 
    $\gamma=0.10,0.40,0.80,1.00$ for the Ising spin case.
    All curves are plotted with one standard error.}
    \label{fig7}
\end{figure*}

We do not give much attention to the spin structure factors 
in the high-temperature paramagnetic state. In such a regime,   
the key magnetic property is submerged by the strong thermal 
fluctuations. At the infinite temperature limit, this is totally true 
but at not high temperature, thermal fluctuation may not erase all information.
Comparing Fig.~\ref{fig5}(a$_1$) with Fig.~\ref{fig5}(d$_1$)(g$_1$), 
they are obviously different, though they are all in the paramagnetic phase 
with the same temperature ${T_1=3J_1}$. This difference originates from 
different values of $\gamma$. In the above context, we know that 
at zero temperature ${\gamma<\gamma_c}$ 
and ${\gamma>\gamma_c}$ drive the system into the N\'eel state and 
the spiral spin liquid, respectively. For ${\gamma=0.08}$, 
the spin configurations of or near the N\'eel order have lower energies than 
other configurations. At high temperatures, 
the peaks of spin structure factors will broaden around the $\Gamma$ point.
For other $\gamma$ values, the peaks of the spin structure factors for the ground 
state will broaden at high temperatures as well. Qualitatively, 
Fig.~\ref{fig5}(d$_1$) and Fig.~\ref{fig5}(g$_1$) are 
the results of spiral contours in Fig.~\ref{fig5}(f$_1$) 
and Fig.~\ref{fig5}(i$_1$) after thermal broadening, respectively. 
Although most information about the low-temperature property is already 
drowned in the thermal fluctuation, the difference between Fig.~\ref{fig5}(a$_1$) 
and Fig.~\ref{fig5}(d$_1$)(g$_1$) still can tell us the spin structure factors 
rough shape at low temperatures. That is to say, if a similar structure, 
such as that in Fig.~\ref{fig5}(a$_1$), is measured in 
experiment, it is enough to exclude the probability of the SSLs.

\section{Persistent reciprocal ``kagom\'e'' structure for Ising spin}
\label{sec3}

For the Ising spin, one cannot construct the (non-collinear) 
spin spirals with the Ising moments. As we demonstrate below, however, 
there exists an emergent ``kagom\'e'' structure of the spin 
correlation in the reciprocal space due to the Ising nature of 
the local moment and the frustration. This reciprocal  ``kagom\'e'' 
structure is reminiscent of the spiral contour at ${\gamma=0.5}$ 
for the continuous spins. In addition, we show that this structure is 
persistent for the Ising spins as long as ${\gamma>1/4}$. 
We attribute this phenomenon to the ``stiffness'' of the Ising spin.

To begin with, we first rely on the geometric method to obtain the 
lowest energy and the constraint of ground-state spin configurations
for the Ising spin. Due to the discrete nature of the Ising spin, 
the upper limit of $\gamma$ for the N\'eel order is no longer $1/6$, 
which is the value for the continuous spins. The spin configuration of 
the N\'eel order is the same as the continuous 
spin such that $\bfS_1, \bfS_2$, and $\bfS_3$ are all antiparallel to $\bfS_0$ 
with only the global $\mathbb{Z}_2$ symmetry (see the cluster unit in Fig.~\ref{fig1}). 
If flipping one of $\bfS_1, \bfS_2$, or $\bfS_3$ reduces the total energy 
compared with the N\'eel state, this indicates that, the N\'eel order 
is no longer the ground state, and the critical value of the N\'eel 
order is found to be ${\gamma_c=1/4}$. 
When ${\gamma>\gamma_c}$, the ground state is massively 
degenerate as shown in Fig.~\ref{fig1}, and the local constraint 
is two up spins and two down spins on each unit $\triline$. 
Remarkably, we find that this constraint stays the same and valid for any 
${\gamma>\gamma_c}$, and this means that the spin structure factors 
of the ground state for different $\gamma$'s share similar structures. 
This is fundamentally different from the case for the continuous spin 
where the spiral contour varies with the parameter $\gamma$. 
In the following, we perform the numerical simulation
and provide the theoretical understanding. 
It is found that, the ``stiffness'' of the Ising spin 
pins the spin structure factors at low temperatures 
to a reciprocal ``kagom\'{e}'' structure in the momentum space. 
Moreover, the $\mathbb{Z}_2$ symmetry will not lead to the chiral symmetry 
and the chiral phase transition does not exist for any $\gamma$.

\begin{figure}[t]
    \includegraphics[width=1.0\linewidth]{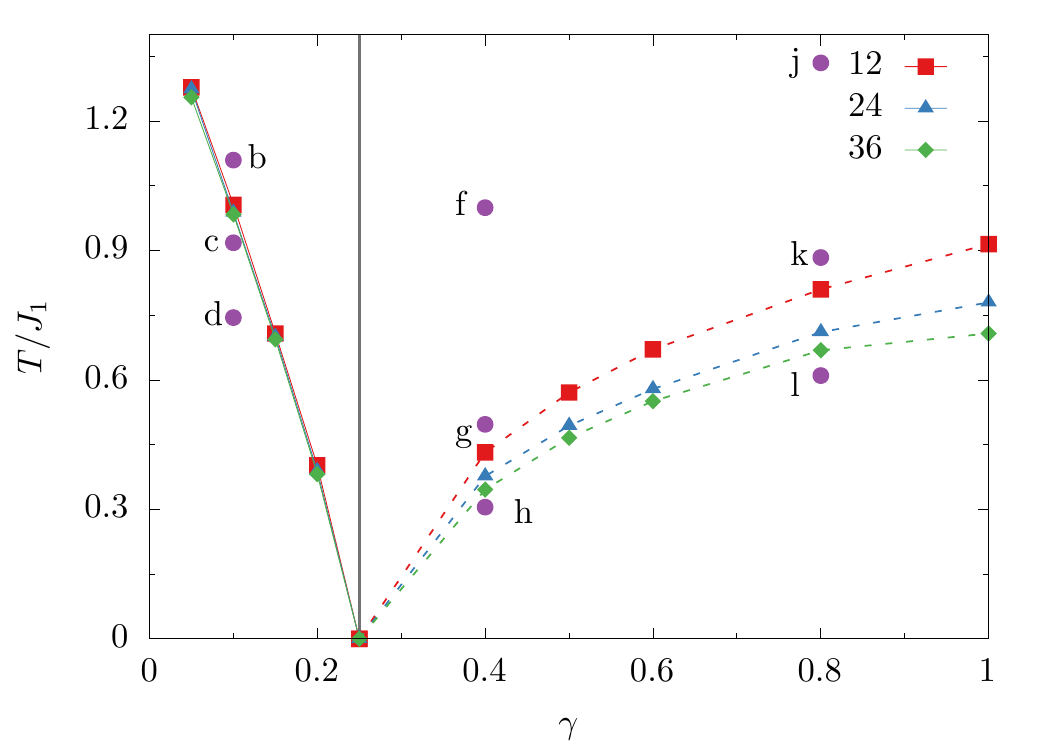}
    \caption{The phase diagram of the Ising spin. The vertical line indicates
     ${\gamma_c=1/4}$ separating the frustration-free and frustrated regions. 
     For ${\gamma<\gamma_c}$, there is a N\'eel-paramagnetism phase 
     transition. For ${\gamma>\gamma_c}$, we choose the low-temperature    
     crossover. The spin structure factors for the purple points are plotted 
     in Fig.~\ref{fig9}. }
    \label{fig8}
\end{figure}

Employing the same Monte Carlo simulations but without invoking
the over-relaxation method, which is invalid for discrete spin, 
we show the specific heat $\Cv$ at 
${\gamma=0.10,0.40,0.80}$ in Fig.~\ref{fig7}. 
The finite-temperature phase diagram 
can be obtained as shown in Fig.~\ref{fig8} where the ``boundary''
is indicated by the peaks of the specific heat.
When ${\gamma<\gamma_c}$, the phase boundary is the 
N\'eel-paramagnetic phase transition where the specific heat 
is divergent accompanied by spontaneous $\mathbb{Z}_2$ 
symmetry breaking. In spite of there being no theorem to restricting the occurrence of
the long-range orders, there is only crossover at finite temperatures 
for ${\gamma>\gamma_c}$ due to strong frustration. 
The sharp peaks in the specific heat occur at lower temperature for larger 
system sizes, and only round maxima are expected to occur
at high temperatures for the thermodynamic limit with ${L\rightarrow\infty}$.
The phase diagram in Fig.~\ref{fig8} shows that the crossover temperatures 
increase with $\gamma$ as at this time the more important interaction 
comes from the exchange $J_2$. The round maximum shown with the dashed curves 
of Fig.~\ref{fig8} is a crossover separating the less correlated higher-temperature 
regime and more correlated low-temperature regime where the reciprocal kagom\'{e} 
regime becomes more and more visible. In highly frustrated magnets, a round maximum 
at the finite temperature of the specific heat is quite common and often observed.
It is simply an indication of the large entropy loss at the crossover temperature point.

Previous works on the same Ising model also explored the thermodynamic properties. 
In Ref.~\onlinecite{Acevedo2021phase}, the authors gave a very similar phase diagram  
of the model via unsupervised machine learning with the system size fixed to 900
sites based on training data obtained from Monte Carlo simulations 
with the Metropolis update and thermal annealing. In addition,  the authors of
Refs.~\onlinecite{Zukovic2021critical,Schmidt2021} did interesting work
on the same model with some focus on the parameter region with 
${\gamma<\gamma_c}$ where the ground state is a simple N\'eel state. 
Through finite-size scaling analysis, they found that the phase transition 
belongs to the 2D Ising universality class for ${\gamma<0.2}$ 
but remains unknown for ${0.20<\gamma<\gamma_c}$ 
due to the rapidly increasing autocorrelation times. 
The authors of Ref.~\onlinecite{Zukovic2021critical} further analyzed the 
dynamical behaviors of the thermal annealing method and 
the parallel tempering with the Metropolis update. 
They found that the low-temperature spin configurations obtained 
from thermal annealing are not always the correct state,
and this might be the reason for the quantitative difference 
between Ref.~\onlinecite{Acevedo2021phase} and our results.

\begin{figure}[t]
    \includegraphics[width=1.0\linewidth]{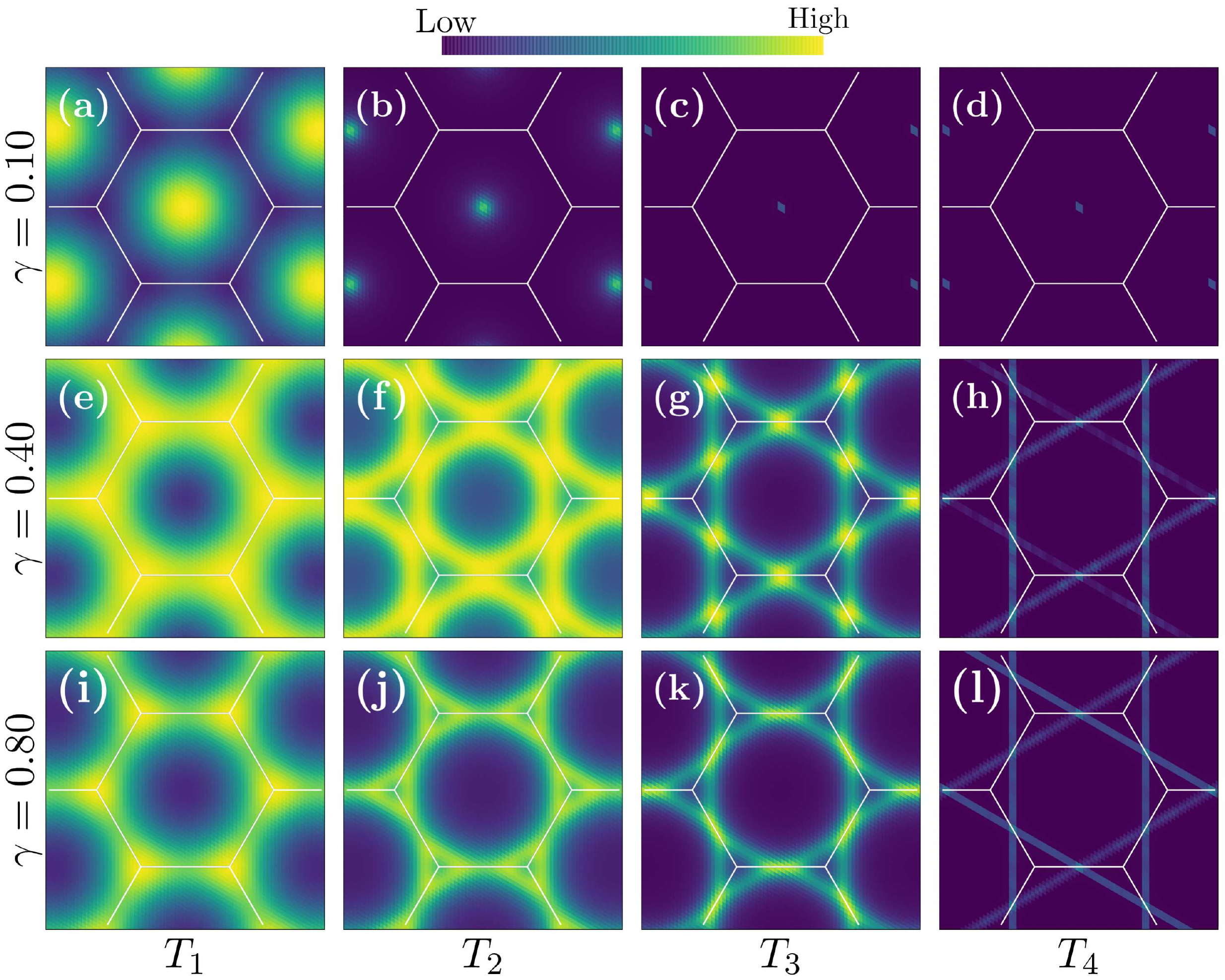}
    \caption{Spin structure factors of the Ising spins at 
    ${\gamma=0.10}$, $0.40$, $0.80$. Temperature 
    ${T_1(3J_1>T_2>T_3>T_4)}$, $T_2$, $T_3$, and $T_4$ 
    are marked in Fig.~\ref{fig8} with purple points. 
    In (a,e,i), 
     the strength colored by the deep 
     purple is non-zero and it is zero in other figures.
     The intensity forms a reciprocal ``kagom\'{e}'' 
     structure for all three $\gamma$'s here.}
    \label{fig9}
\end{figure}

Thermodynamic results such as the specific heat provide rather limited information
about the physical properties of the system at low temperatures. We further 
seek to understand the spin correlations or the spin structure factors that would   
give more important characterization of the low-temperature physical properties. 
These results bring us some understanding of the Ising-spin-based 
frustrated magnetism. In Fig.~\ref{fig9}, we depict the (Ising) spin structure factors 
for different temperatures and different $\gamma$'s. 
As long as ${\gamma>\gamma_c=1/4}$, the spin structure factors develop a reciprocal 
``kagom\'e'' structure in the momentum space 
at low temperatures. 
This reciprocal ``kagom\'e'' structure can be clearly observed in
Fig.~\ref{fig9}(h) and Fig.~\ref{fig9}(l) for ${\gamma=0.40}$
and ${\gamma=0.80}$, respectively. 
This contrasts strongly with the case of continuous spins 
where the reciprocal ``kagom\'e'' structure only occurs for ${\gamma=0.5}$
%\CJH{\sout{and does not occur elsewhere (see Fig.~\ref{fig1} and Fig.~\ref{fig5}).}
and does not occur elsewhere (see Figs.~\ref{fig2}, \ref{fig5} and \ref{fig6}).

\begin{figure}[b]
    \includegraphics[width=0.78\linewidth]{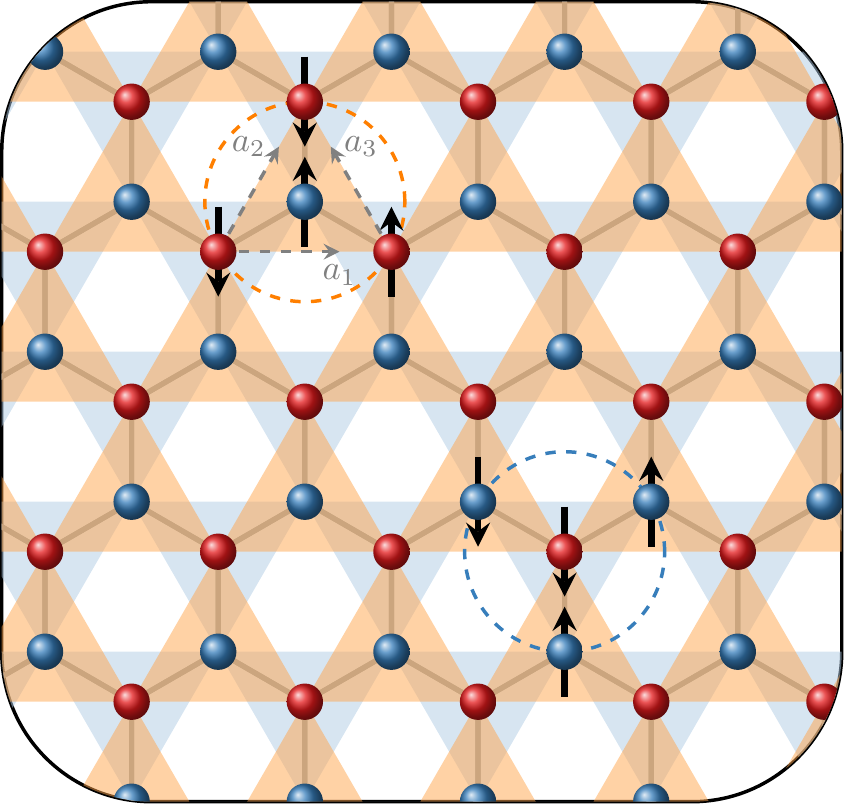}
    \caption{Honeycomb lattice with units in orange and blue colors. Here
   ${\bf a}_{\mu}$  ($\mu=1,2,3$) are the Bravais lattice vectors. }
    \label{fig10}
\end{figure}

\begin{figure*}[t]
    \includegraphics[width=\textwidth]{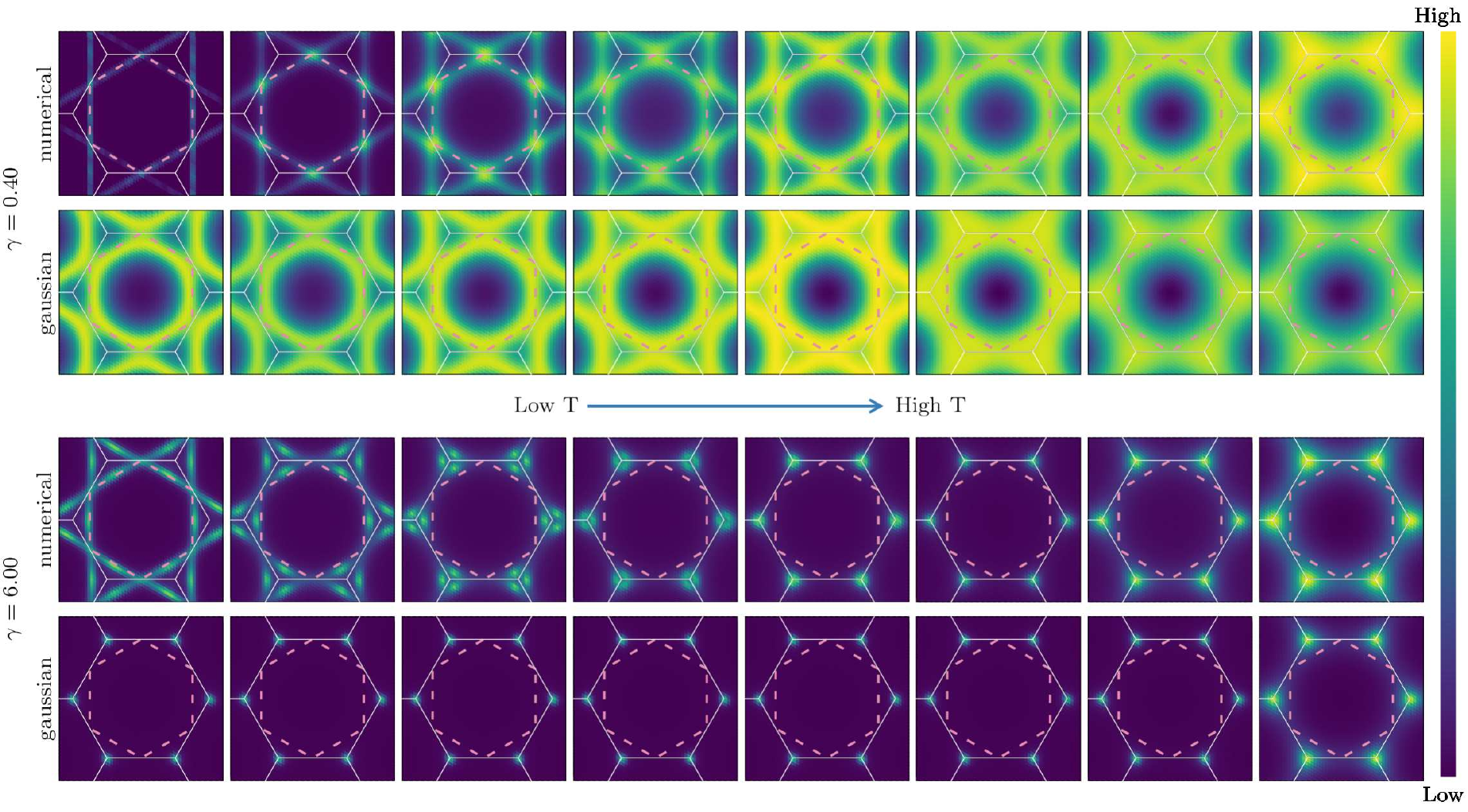}
    \caption{Evolution of the spin structure factors for the Ising 
    spin with ${\gamma =0.4}$ and ${\gamma =6.0}$ 
    as a function of temperature with a comparison with the 
    results from the self-consistent Gaussian approximation. 
    Two animations of the numerical results of the thermal
    evolution can be found in the Supplemental Material~\cite{Supple}. 
    The dashed inner hexagon is for guidance. 
    The intensity bar indicates the relative intensity for each panel. 
    In the upper set of panels, the temperatures from left to right are 
0.306, 0.362, 0.471, 0.624, 0.862, 1.279, 2.086, and 4.000
in units of $J_1$. In the lower set of panels, the temperatures 
from left to right are 0.446, 0.556, 0.663, 0.728, 0.778, 1.279, 
2.086, and 4.000 in units of $J_1$. 
    }
    \label{fig11}
\end{figure*}

To understand the persistent reciprocal ``kagom\'e'' structure for 
the Ising spin with  ${\gamma>\gamma_c}$, we return to the geometric 
method in Sec.~\ref{sec2a}. The constraint in Eq.~\ref{eq7}
cannot be satisfied for the Ising spin. To
optimize the energy, one essentially minimizes
 $S_0 + 2\gamma S_{\mytriangle}$. This then requires 
the three spins, $S_1, S_2$, and $S_3$, on the triangular corner 
of the unit ``\triline'' to have either ``two-up-one-down'' or 
``two-down one-up'' configuration, 
and the central spin, $S_0$, to align with the minority spin. 
For example, for the blue triangular sublattice in Fig.~\ref{fig10},
only the blue triangles have  two-up-one-down or 
two-down-one-up spin configuration, not the white region. 
A similar requirement applies to the orange triangular sublattice. 
These requirements actually differ from the energy optimization 
for the triangular lattice Ising antiferromagnet that demands 
all the triangles to be either two-up-one-down or 
two-down-one-up. The local constraint implies that
a pair of the Ising spins separated by ${\bf a}_{\mu}$ (${\mu=1,2,3}$) 
is mostly anti-collinearly aligned. At low temperatures, the spins 
are fluctuating near the ground state manifold. This means that 
the low-temperature spin structure factor in the momentum space 
would be peaked at ${{\bf k}\cdot {\bf a}_{\mu}= \pm \pi}$. 
These peaked momenta form a reciprocal ``kagom\'{e}'' structure. 
One specific construction is given here. 
If one chooses ${{\bf k}\cdot {\bf a}_{1}= \pm \pi}$
that fixes the $x$ components of the peaked momenta, then the Ising 
spins are anti-collinearly aligned along the ${\bf a}_1$ direction. 
Once one ${\bf a}_1$-directed spin chain with anti-collinearly aligned
Ising spin is realized, the two-up-one-down or 
two-down-one-up condition is satisfied automatically on all the 
triangular units boarding the spin chain. Thus, the neighboring
${\bf a}_1$-directed spin chains are not constrained except being 
anticollinearly aligned along ${\bf a}_1$. Thus the $y$ component
of the peaked momenta can be arbitrary. The other equivalent 
momenta can be generated by the crystal symmetry of the system.

\section{Thermal evolution of reciprocal structure for Ising spins} 
\label{sec4}

In the previous section, we have shown and argued that 
the discrete nature of the Ising moment and the frustration 
lead to an unexpected reciprocal ``kagom\'{e}'' structure 
in the spin structure factors at low temperatures. Moreover,
this reciprocal structure is persistent with the varying of the 
exchange parameters. 
On the other hand, in several early publications, it was 
shown that the self-consistent Gaussian approximation 
seems to work even for frustrated Ising 
antiferromagnets~\cite{PhysRevB.94.224413,PhysRevB.92.220417,
PhysRevLett.93.167204}. 
This method was known to work well for
frustrated XY or Heisenberg antiferromagnets. As we have found that the 
Ising spin behaves rather differently from the XY or Heisenberg spins 
at low temperatures, it is then hard to expect that 
the self-consistent Gaussian approximation will continue to 
work well for the Ising model. Apparently, 
if one directly applies the self-consistent Gaussian 
approximation to evaluate the spin structure factors, 
the reciprocal ``kagom\'{e}'' structure cannot be obtained, and the 
reciprocal contours would be more like the ones for the 
continuous spins. Then how do we reconcile the persistence
of the reciprocal ``kagom\'{e}'' structure and the presumed 
applicability of the self-consistent Gaussian approximation? 
In addition, what is the fate of the persistent reciprocal 
``kagom\'{e}''s structure in the presence of the thermal fluctuations? 
This is addressed below.

We here fix the $\gamma$ values and evaluate the spin 
structure factor for the Ising spins by varying the temperature. 
For convenience of comparison and presentation, 
we display the results from low temperatures to
high temperatures in Fig.~\ref{fig11}. The animations of these results 
can be found in the Supplemental Material~\cite{Supple}. 
We find that, at low temperatures, the spin structure factor
is peaked at the reciprocal ``kagom\'{e}'' structure. 
As the temperature increases, the peak position 
of the spin structure factor 
gradually deviates from the reciprocal ``kagom\'{e}'' structure. 
In the high temperature limit, one cannot trace any signature 
of the reciprocal ``kagom\'{e}'' structure. 
This is expected due to the thermal fluctuations. 
At high temperatures, the spins fluctuate strongly,
and the system deviates from the ground state manifold. 
Hence the stiffness of the Ising spin, which is partly responsible 
for the reciprocal ``kagom\'{e}'' structure, is conquered by thermal
fluctuations.

We then perform a self-consistent Gaussian approximation
to calculate the spin structure factors at finite temperatures.  
In this scheme, we first write down the partition function for the
system,
\begin{eqnarray}
{\mathcal Z} &=& \int {\mathcal D} [S_i] \,
e^{-\beta \sum_{ij} J_{ij} S_i S_j } \,
\prod_i \,  \delta ( S_i^2 - S^2) 
\nonumber \\
&=& \int {\mathcal D} [S_i]  {\mathcal D} [\lambda_i] \,
e^{-\beta \sum_{ij} J_{ij} S_i S_j + \sum_i \lambda_i (S_i^2 -S^2)  }
\nonumber \\ 
&\equiv & \int {\mathcal D} [S_i]  {\mathcal D} [\lambda_i] \ e^{- 
{\mathcal S}_{\text{eff} } [\beta, \lambda_i]},
\end{eqnarray}
where  $\lambda_i$ is the Lagrange multiplier that imposes 
the magnitude constraint for the Ising spin with ${|S_i|= S }$,
and ${\mathcal S}_{\text{eff} }$ is the effective action and given 
as
\begin{eqnarray}
{\mathcal S}_{\text{eff} } &=& \sum_{\bf k } \beta 
\big[\frac{1}{2} {\mathcal J}_{\mu\nu} ({\bf k})
+ \Delta (T) \delta_{\mu\nu} \big] S^{\mu} ({\bf k }) S^{\nu} (-{\bf k})
\nonumber \\
&& \quad\quad\quad\quad\quad\quad\quad\quad\quad 
- S^2 \beta N \Delta (T). 
\end{eqnarray}
Here we have made a saddle point approximation by setting
${\lambda_i = -\beta \Delta (T)}$, which is equivalent to replacing the 
single-spin constraint with a global spin constraint. 
The uniform saddle point $\lambda$ is based on the property 
of the homogeneous paramagnetic state that respects all the lattice symmetry. 
Thus the spin correlation function is then given as 
\begin{eqnarray}
\langle 
S^{\mu} ({\bf k}) S^{\nu} (-{\bf k})
\rangle 
 = \frac{1}{\beta} 
\big[  
\frac{{\mathcal J} ({\bf k}) }{2} + \Delta (T) {\mathbb I}_{2\times 2}
\big]^{-1}_{\mu\nu} ,
\end{eqnarray}
where $ {\mathbb I}_{2\times 2}$ is a $2\times 2$ identity matrix. The 
saddle point parameter $\Delta (T)$ is determined self-consistently 
from the saddle point equation,
\begin{eqnarray}
\sum_{\bf k} \sum_{\mu } 
\frac{1}{\beta} \big[  \frac{ {\mathcal J} ({\bf k}) }{2} 
+ \Delta (T ) {\mathbb I}_{2\times 2} \big]^{-1}_{\mu\mu} 
= NS^2.
\end{eqnarray}
Clearly, in this scheme, the momentum information 
arises from the exchange interaction matrix ${\mathcal J}({\bf k})$.
Thus, the spin structure factor is mostly weighted around the 
degenerate contours of the lowest eigenvalues of ${\mathcal J}({\bf k})$. 
This is a bit analogous to the spiral spin liquid regime for the continuous
spins in Sec.~\ref{sec2}. In fact, we have computed the spin structure 
factors within the self-consistent Gaussian approximation 
for the continuous spins in Fig.~\ref{fig5}. 
Our results from the Ising spins 
are depicted in Fig.~\ref{fig11}. 
In the upper (lower) set of panels of Fig.~\ref{fig11},
we choose ${\gamma=0.40}$ (${\gamma=6.0}$) 
such that the contour is around the center (corner) of the 
Brillouin zone. The comparison with the numerical results 
is better at the high temperatures and is poorer at the low temperatures.
This behavior is expected. In the high temperatures, the Ising spins are
 strongly fluctuating thermally, and the spin magnitude constraint
does not play a strong role. The system widely leaves the ground-state manifold
such that the reciprocal ``kagom\'{e}'' structure is no longer visible, but 
the correlations between the spins are still preserved from the exchange interaction
part. The self-consistent Gaussian approximation captures this high temperature 
correlated regime and gives a qualitatively reasonable description of the 
spin structure factors. 
As a comparison for the continuous spins in Fig.~\ref{fig5}, 
the spiral contours are well captured by the self-consistent 
Gaussian approximation from the high temperatures to the 
low temperatures except that the intensity is not
well obtained in the low-temperature limit.

% momentum information, not self-consistent gaussian approximation,  the constraint

\section{Discussion}
\label{sec5}

We have shown that the spiral spin liquid regime holds for both XY
and Heisenberg spins in the frustrated regime at low but finite temperatures. 
We further identified the finite-temperature chiral transition for both cases.
 For the Ising spin, a reciprocal  ``kagom\'e'' 
structure emerges in the low-temperature spin structure factors   
in the momentum space, and persists for a range of exchanges 
in the frustrated regime which resembles the spiral spin liquid regime. 
This is understood from the stiffness of the discrete Ising spins and the 
frustration due to the competing interactions on the honeycomb lattice. 
Moreover, the reciprocal structure evolves from the ``kagom\'e'' 
structure into the contours demanded by the soft spin analysis 
as the temperature is increased.

It is readily evident as well as illuminating to notice that, if one adds 
an anisotropic term such as $D (S_i^x)^2$ [or $D (S_i^z)^2$] to the model
for the XY (or Heisenberg) spin, the system will behave like the Ising 
spin with an increasing and positive $D$. Thus one anticipates
that the spiral contour of the XY and Heisenberg cases will crossover 
to the reciprocal kagom\'e structure. The details of such a crossover 
process were not pursued in this paper, and we expect that it could be 
observed in real materials of the spiral spin liquid regime through 
the tuning of such spin anisotropy, e.g. via hydrostatic 
pressure~\cite{Lin2018,Sun2018effects,Sakurai2021}.

\subsection{Experimental connection}

We here discuss the experimental connection. As far as the materials' relevance, 
many honeycomb lattice van der Waals magnets have been proposed and
synthesized. For a recent review of them, one can refer to 
Ref.~\cite{Gong2019two}. More recently, rare-earth honeycomb lattice 
magnets were proposed and studied~\cite{Li_2017,Luo_2020,2019PhRvB..99x1106J}. 
For the rare-earth magnets, the exchange interaction is often anisotropic
and short ranged. The exchange coupling decays rapidly with the distance
and often becomes quite weak beyond the first neighbor. 
Thus, the frustration on the rare-earth honeycomb lattice 
magnets often arises from the anisotropic exchange, rather 
than the competing further-neighbor interactions. 
From this perspective, it seems a bit difficult to expect the physics 
in this paper to be directly relevant for these honeycomb lattice 
rare-earth magnets. In fact, 
the Yb honeycomb lattice magnet YbCl$_3$ was found to have a
simple N\'eel state~\cite{PhysRevB.102.014427}, indicating the 
dominance of the first-neighbor interaction~\cite{PhysRevB.100.180406}. 
However this is not the end of the story for the honeycomb lattice 
rare-earth magnets.

A series of vdW rare-earth chalcohalides were recently 
synthesized~\cite{2021arXiv210312309J}. These materials 
are not of planar honeycomb structure like YbCl$_3$~\cite{PhysRevB.102.014427};
instead, they are formed by the bilayer triangular lattice of the rare-earth moments.
The bilayer triangular lattice is A-B stacked and is equivalent to 
a honeycomb lattice. The first-neighbor and second-neighbor
distances are close. For HoOF, the first neighbor is $3.57$\AA,
and the second neighbor is $3.80$\AA. Some of these compounds
certainly support highly anisotropic spin interactions such as 
the Kitaev interaction~\cite{Luo_2020,2019PhRvB..99x1106J,Li_2017}; 
the Ho-based, Dy-based, and even Er-based ones could
actually provide the Ising local moments~\cite{2021arXiv210312309J}
as the effective moments of these ones are close 
to the fully polarized atomic values. If we assume that
both the first- and second- neighbor Ising interactions are 
due to the dipole-dipole interaction, this will put these materials 
in the frustrated regime with ${\gamma > \gamma_c}$. 
Thus, if the remaining weak interactions beyond the first and 
second neighbor Ising interactions do not play a significant role, 
we expect our prediction including the reciprocal ``kagom\'{e}'' structure
and the thermal evolution of the reciprocal structure in this paper 
to be applied to these compounds. Experimentally, this requires 
an inelastic neutron scattering measurement of the dynamic
spin correlation for a range of energies that is integrated 
to yield the equal-time spin correlation.

In this family of materials,  the Yb and Sm ones seem to be more quantum,
and most likely to carry quantum spin-1/2 Kramers doublets. According to
a more microscopic study of the exchange paths~\cite{PhysRevB.98.054408},
the exchange interaction between the Yb local moments may 
sometimes be Heisenberg-like. In that case, the Yb compound 
may realize a competing spin-1/2 $J_1$-$J_2$ Heisenberg model,
and the Dzyaloshinskii-Moriya interaction could play some role here. 
The spiral spin liquid regime may apply to the finite temperature regime 
of the spin-1/2 $J_1$-$J_2$ Heisenberg model, but the ground state 
of this model can be interesting as well~\cite{PhysRevLett.110.127203}.
In addition, in the $J_1$-$J_2$ model with the Heisenberg spin, 
adding an external field perpendicular to the lattice could bring 
about a skyrmion crystal~\cite{Muhlbauer2009skyrmion,Okubo2012multiple},
and thermal Hall effect of magnons could be generated.  
The magnetic skyrmion was suggested to have a great stability on account of 
its topological protection, which makes it an excellent information carrier, 
and its topological properties induce a variety of emergent behaviors 
attracting research interest. In the vdW materials, this will not only 
have theoretical value but also promote the development of device 
applications.

The Co-based honeycomb lattice antiferromagnets were recently 
proposed to be candidates for Kitaev materials~\cite{PhysRevB.97.014408,
PhysRevB.97.014407} due to the spin-orbit-entangled local moment. 
In fact, the Co-ions were also known to support Ising interactions. 
This has been found in the quasi-one-dimensional Ising magnets 
CoNb$_2$O$_6$, BaCo$_2$V$_2$O$_8$ and CaCo$_2$V$_2$O$_8$~\cite{Coldea2010,PhysRevLett.123.067202,PhysRevB.91.140404,Wang2018}. 
It would also be interesting to search for Co-based honeycomb lattice antiferromagnets that realize Ising spins. 

Another set of compounds that are not really magnets and do not rely on the 
spin degrees of freedom are the Fe-based mixed valence compounds. 
The system contains both Fe$^{2+}$ ions and Fe$^{3+}$ ions, which can be 
equivalently treated as Ising spin, and thus realizes an interacting Ising spin system.
This idea has been applied to understand the electron charge 
physics in LuFe$_2$O$_4$ and YbFe$_2$O$_4$~\cite{PhysRevB.94.224413,PhysRevB.92.220417}.
The lattice in these two compounds 
is not honeycomb but frustrated. If an Fe-based mixed valence compound 
with a honeycomb lattice is found, our result can be applied, and the spin 
correlation should be replaced by the electron density-density correlation,
which can be detected by X-ray scattering.

\subsection{Frustrated Ising magnets }

% comment on Gaussian approximation 

Here, we discuss the implications of our results for frustrated Ising magnets. 
In this paper, we show that the high-temperature spiral spin liquid regime is not 
quite sensitive to the spin dimensions. For both the continuous spin and the 
Ising spin, the same kind of momentum-space contours demanded by the 
exchange interaction appear in the spin structure factor. 
This is well captured by the self-consistent Gaussian approximation.
However, at low temperatures, the Ising spin in the honeycomb lattice of our problem 
develops an unconventional reciprocal ``kagom\'{e}'' structure. 
This means that frustrated Ising magnets may contain more interacting
correlated spin structures at low temperatures than the ones expected from
the self-consistent Gaussian approximation. 
Frustrated Ising magnets in an A-B stacking multilayer triangular lattice
or an A-B-C stacking multilayer triangular lattice, which were studied in 
LuFe$_2$O$_4$ and YbFe$_2$O$_4$~\cite{PhysRevB.94.224413,PhysRevB.92.220417}, may contain some new ingredients at low temperatures
beyond what has been found using 
the high-temperature Gaussian approximation, and may be worth 
a careful examination. A similar situation may also occur for the 
diamond lattice Ising antiferromagnet in the frustrated regime.

\section*{Acknowledgments}

C.-J. H. and J.Q.L. thank Xu-Ping Yao and Changle Liu for discussions. 
GC is thankful for the hospitality of Long Zhang and Fuchun Zhang at Kavli Institute
of Theoretical Sciences where this work was completed. 
The numerical part of this work was performed on TianHe-2. We are thankful for 
the support from the National Supercomputing Center in Guangzhou (NSCC-GZ).
This work is supported by 
 the Ministry of Science and Technology of China 
with Grants No.~2018YFE0103200, 2021YFA1400300, the National Science Foundation of China with 
Grant No.~92065203,
by the Shanghai Municipal Science and Technology Major Project with 
Grant No. 2019SHZDZX01, and by the Research Grants Council of Hong Kong 
with General Research Fund Grant No. 17306520.

%\bibliography{refs}
%\input{publish.bbl}


\end{document}